\documentclass[prb,twocolumn,superscriptaddress,numerical,longbibliography]{revtex4-2}
\usepackage[utf8]{inputenc}
\usepackage{graphicx}
\usepackage{amsmath}
\usepackage{braket}
\usepackage{ytableau}
\usepackage{amssymb}

\begin{document}
\title{Characteristic features of the strongly-correlated regime: Lessons from a 3-fermion one-dimensional harmonic trap}

\author{Victor Caliva}
\affiliation{Universidad de Buenos Aires, 
Facultad de Ciencias Exactas y Naturales,
Departamento de F\'isica. Buenos Aires, Argentina}
\affiliation{CONICET - Universidad de Buenos Aires, 
Instituto de F\'isica de Buenos Aires (IFIBA). Buenos Aires, Argentina}

\author{Johanna I.\ Fuks}
\affiliation{Universidad de Buenos Aires, 
Facultad de Ciencias Exactas y Naturales,
Departamento de F\'isica. Buenos Aires, Argentina}
\affiliation{CONICET - Universidad de Buenos Aires, 
Instituto de F\'isica de Buenos Aires (IFIBA). Buenos Aires, Argentina}

\date{\today}

\begin{abstract}
 The transition into a strongly-correlated regime of 3 fermions trapped in a one-dimensional harmonic potential is investigated.
This interesting, but little-studied system, allows us to identify characteristic features of the regime,
some of which are also present in strongly-correlated materials relevant to the industry.  
Furthermore, our findings describe the behavior of electrons in quantum dots, ions in Paul traps, and even fermionic atoms in one-dimensional optical lattices. Near the ground state, all these platforms can be described as fermions trapped in a harmonic potential. 
The correlation regime can be controlled by varying the natural frequency of the trapping potential, and to probe it, we propose to use twisted light.
We identify 4 signatures of strong correlation in the one-dimensional 3-fermion trap, which are likely to be present for any number $N$ of trapped fermions:
i) the ground state density is strongly localized with $N$ maximally separated peaks (Wigner Crystal) 
ii) the symmetric and antisymmetric ground state wavefunctions become degenerate (bosonization) iii) the von Neumann entropy grows,
iv) the energy spectrum is fully characterized by $N$ normal modes or less.
\end{abstract}

\maketitle

\section{Introduction}
Forty years ago Feynman envisioned quantum emulators (also known as analogue quantum computers) in his famous statement: “Nature isn't classical, dammit, and if you want to make a simulation of nature, you'd better make it quantum mechanical, and by golly it's a wonderful problem, because it doesn't look so easy” \cite{Feynman}.
An analogue quantum computer is a controllable quantum system able to simulate the behavior of another, more challenging (less controllable, larger, more complex) quantum system \cite{Kendon10, AnalogQComp23, Qemulators23}. Such a quantum emulator is not universal, and unlike gate quantum computers, it can not be programmed and there is no translation of an algorithm into a quantum circuit. Therefore, to simulate a given problem, one needs to devise an experimentally feasible and controllable quantum system able to emulate the behaviour of the original problem. 
Several studies show the potential of noisy cold atoms and ion traps as analogue quantum computers \cite{IonSimu, Qsimulation, adv_5, Cirac22}.
Quantum advantage has been demonstrated for the Gaussian boson sampling problem using a photon quantum emulator\cite{adv}.

Strong correlation is particularly challenging and has been widely studied, both in-silico and experimentally  \cite{raman,st,st_2,st_3,st_4,Serwane}. 
Stretched molecular bonds involved in biochemical processes and transition-metal catalysts for energy storage devices are some prominent examples of strongly-correlated problems that are relevant to the industry. 
In the strongly-correlated limit some materials 
exhibit novel phenomena not observed in weakly correlated materials. Some examples are metal-insulator transitions \cite{Feng21}, magnetic order, unconventional superconductivity
\cite{Adler18}, superfluidity, quantum phase transitions, and topological order \cite{geo, tar}.
Designing and optimizing materials
with these properties is key to advancing technology 
and poses a great challenge for the simulation community.
Hartree Fock is a poor solution for these types of problems and DFT, which is very successful in predicting properties of weakly correlated materials, requires sophisticated exchange-correlation functionals to achieve qualitatively correct results in this regime \cite{Cohen12, Graefenstein04, Kronik22}. Exchange-correlation functionals designed for strong correlation, on the other hand, do not perform well for more weakly correlated materials \cite{Malet12}. More expensive wavefunction methods such as CCSD also struggle to simulate strong correlation \cite{Scuseria18} and Complete Active Space approaches require large active spaces to converge \cite{Marti08}. The strongly-correlated regime of quantum many-body systems is therefore an interesting application for quantum computing. 

In this work, we study one-dimensional harmonic traps of trapped fermions as a testbed to learn about the characteristic features of the strongly-correlated regime in fermionic many-body systems. 
The 2-fermion one-dimensional harmonic trap was previously studied in Ref.~\cite{ic1/ic2}. We generalize the results to 3 fermions and infer the behaviour of $N$-trapped fermions. Analytical results with cold fermionic atoms trapped in optical lattices have revealed that phenomena previously investigated in the presence of many atoms may be studied in the limit of a few particles as well \cite{Loft, Cioslowski, Balzer, ions, damico}. 
Near the ground state, electrons and holes in quantum
wires and dots, ions in Paul traps, and cold fermionic atoms in
optical lattices are all realizations of the fermion harmonic trap \cite{Yip, ions_2, sar, blatt}. 
Besides the importance of quantum dots in the display industry, and of ion traps in the development of digital quantum computers  \cite{Brown21}, these platforms can all serve as analogue quantum computers, also known as quantum emulators.
For atoms, the interaction is modeled as a contact potential \cite{cold_atoms}, whereas for electrons and ions the interaction is via a Coulomb potential \cite{Brey,Cou_2} or soft-Coulomb \cite{plata, soft_2} for one-dimensional traps. The confining potential is parabolic and its strength is characterized by the natural frequency $\omega_0$. The correlation
regime in harmonic traps can be controlled by varying
the particle-particle interaction strength or the confinement strength of the parabolic potential $\omega_0$, and it also depends on the number of trapped
particles. 
By virtue of the high tunability and the availability of
analytic (weak and strong interaction limits) \cite{cold_atoms, Taut, Taut_2}
and very accurate one-dimensional numerical solutions
\cite{gha, Tomasz}, these systems provide a means to understand the physics and interpret the experiments on more complex quantum many-body systems. 
The problem of fermions trapped in a parabolic potential has been studied theoretically for 3 electrons in a three-dimensional trap in the limits of weak and strong confinement \cite{Taut}, in the context of electrons \cite{Loft} \cite{Cioslowski}, for 2 electrons in one and two dimensions \cite{Balzer}, for ions \cite{ions} and for fermionic atoms \cite{damico}.

In this work, we derive expressions for the density and normal modes of 3 fermions in a one-dimensional harmonic trap in the strongly correlated limit and use them to validate the numerical simulations. We then perturb the system and analyze the transition into a strongly-correlated regime in the absorption spectrum.

The paper is organized as follows: in section \ref{sec:sep} we discuss the separability of the harmonic trap hamiltonian into center of mass (CM) and internal degrees of freedom and the antisymmetrization of the total wavefunction, in section \ref{sec:analytic} we study the degeneration of the ground state and derive the normal modes and an expression for the spatial wavefunction and the ground state density of the 3-fermion harmonic trap in the strongly correlated limit $\omega_0 \to 0$. In section \ref{sec:numerical} we validate the finite $\omega_0$ numerical simulations against the analytic results derived for the strongly correlated limit: we confirmed 
the ground state density is localized as expected and the energy spectrum contains the predicted frequencies in this limit. We then analyze the quadrupole spectrum obtained from the numerical time-evolution of the density after perturbation with a quadrupole field and discuss the selection rules. In section \ref{sec:exp} we discuss an experimental setup to probe correlation effects in the quadrupole spectrum using twisted light and in section \ref{sec:conclusions} we conclude.

\section{Separability into center-of-mass (CM) and internal coordinates}
\label{sec:sep}

The Hamiltonian describing a harmonic trap with $N$ particles is separable into a center-of-mass (CM) term and a term dependent on the internal degrees of freedom of the system. The CM system is equivalent to a particle with mass $M=Nm$ and charge $Ne$,

\begin{equation} \label{eq:Hcm}
{H_{0}}^{CM}=\frac{{P_{X}}^{2}}{2M} + \frac{1}{2}M{{\omega}_{0}}^{2}{X}^{2}
\end{equation}
\noindent 
with $X=\sum_{i}^{N} x_{i}$ y $P_{X}=\sum_{i}^{N} p_{i}$. For $N=3$ and defining the Jacobi coordinates   as

\begin{equation} \label{eq:q1q2}
q_{1}=\frac{x_{1}-x_{2}}{\sqrt{2}}, q_{2}=\frac{x_{1}+x_{2}}{\sqrt{6}}-\sqrt{\frac{2}{3}}x_{3} ,
\end{equation}

the internal Hamiltonian reads,
\begin{equation} \label{eq:Hq}
{H_{0}}^{q}=\frac{{P_{1}}^{2}}{2m} + \frac{{P_{2}}^{2}}{2m} + \frac{1}{2}M{{\omega}_{0}}^{2}{q_1}^{2} +
\frac{1}{2}M{{\omega}_{0}}^{2}{q_2}^{2} + V(q_1,q_2).
\end{equation}
\noindent 
where $ V(q_1,q_2)$ is the particle-particle interaction potential which is the Coulomb potential \cite{Brey} \cite{Cou_2} $ V(q_1,q_2)=-1/|q_1-q_2|$ in the case of quantum dots and ion traps \footnote{Rydberg atoms are also represented by this Hamiltonian with $V(q_1,q_2)\propto 1/(q_1-q_2)^6$ \cite{cold_atoms}.}. 
 Atomic units,  $ \hbar=e=m_e=a_0=1$, are used throughout the work.
 
Notice that due to the separability of the Hamiltonian the total wavefunction $\Phi(x_1,x_2,x_3)$ can be written as a product of a wavefunction associated with the internal coordinates  $\xi(q_1,q_2)$ and another associated with the  CM coordinates $\zeta(X)$:
\begin{equation} \label{eq:wavef}
\Phi(x_1,x_2,x_3)=\xi(q_1,q_2)\zeta(X). 
\end{equation}
\noindent

\subsection{Fermionic wave function}

In order for the total wave function Eq.~(\ref{eq:wavef}) to represent fermions it must be antisymmetric. The symmetry of the total wave function is obtained from the product of a spin and a spatial function,

\begin{equation} 
\Psi(x,\sigma)=\Lambda(\sigma)\Phi(x).
\end{equation}
\noindent 

To characterize the symmetries of the spin wave function $\Lambda(\sigma)$ we use the concept of Young diagrams, discussed in some depth in Appendix \ref{sec:y}.
For the 3-fermion system, we obtain 8 spin states: 4 are pure symmetric, 2 are symmetric mixed, and 2 are antisymmetric mixed. There are no pure antisymmetric states for this problem \cite{Taut}.

The symmetries that define the spatial wave function $\Phi(x)$ are the Pauli principle and the Parity
\cite{Loft},
\begin{equation} \label{eqpauli}
\Phi(-q_1,q_2,X)=-\Phi(q_1,q_2,X)\:\textup{( Pauli principle)} 
\end{equation}
\begin{equation}\label{parity}
\Phi(-q_1,-q_2,X)=\pm \Phi(q_1,q_2,X)\:\textup{(Parity)}
\end{equation}

The total wave functions $\Psi$ must be assembled \cite{Taut}:

\begin{equation} \label{ant}
{\Psi}_i(1,2,3)=\hat{{\textit{A}}_a}{\Lambda}_i(1,2,3)\Phi(1,2,3)
\end{equation}
\noindent 
with the antisymmetrizer $\hat{{\textit{A}}_a}=\frac{1}{\sqrt{N!}}\sum_{\textit{P}}(-1)^p\hat{\textit{P}}$, and the permutation operator $\hat{\textit{P}}$ \cite{Taut,Ma}. ${\Lambda}_i$ is the spin function, $i$ is equal to 1 for $S=3/2$ in pure symmetric case and $S=1/2$ for mixed symmetric case, and equal to 2 and $S=1/2$ for mixed antisymmetric case. The labels $(1,2,3)$ are spin or spatial coordinates depending on the function in question. Spin functions are discussed in the appendix.

\section{Analytic solution in the strong correlation limit
} 
\label{sec:analytic}
\subsection{Degeneration of the ground state}

In the strong correlation limit, when particle-particle repulsion dominates over parabolic confinement, the two-particle problem has two degenerate ground states with opposite symmetries in the spatial part of the wavefunction.
A solution differs from another solution only by a sign. There are odd functions
and even functions in the spatial part, with their corresponding change in the symmetry of the
spin part. The same procedure can be applied to larger systems, $N > 2$.
For each pair of particles with opposite spins, it is possible to change the symmetry of the spatial part
of the wave function and build new solutions in the limit
of infinitely strong interaction. In this limit, there are multiple degenerate ground states
$D$, the value of $D$ is given by the number of different spin configurations, counting
the different possibilities of dividing $N$ particles between the particles with spin-up $N_{\uparrow}$ and
spin-down $N_{\downarrow}$ \cite{Sowinski}.

\begin{equation} \label{Dconf}
D=\frac{N!}{N_{\uparrow}!N_{\downarrow}!}
\end{equation}

For a system with $N_{\uparrow}=1$ and $N_{\downarrow}=2$, or vice versa (the account of $D$ contemplates the two
configurations) we have $D = 3$, i.e. three wavefunctions with equal energy value.

\subsection{Normal modes}

In the work of Balzer \cite{Balzer} the normal modes of the one-dimensional 2-fermion harmonic trap in the strong correlation limit are derived. The assumption is that the wavefuntion is highly localized at $X_0$ and $-X_0$ in such limit  and the solution is found using a first order Taylor expansion around these equilibrium positions. 
In this work we generalized Balzer's approach for the case of 3 fermions to find the charactersitic frequencies (normal modes) of the system. The potential in the original coordinates reads,

\begin{equation} 
\begin{split}
V(x_1,x_2,x_3)= \frac{1}{2}{{\omega}_{0}}^{2}({x_1}^{2}+{x_2}^{2}+{x_3}^{2})
\\ +\frac{1}{|x_1-x_2|}+\frac{1}{|x_2-x_3|}+\frac{1}{|x_1-x_3|}
\label{Balzer3e}
\end{split}
\end{equation}
\noindent
where $x_1$, $x_2$ y $x_3$ are the positions of each of the 3 electrons. We set the gradient of the potential to zero to obtain the minimum equilibrium positions $\overrightarrow{x_0}=(x_1=x_0,x_2=0,x_3=-x_0)$, with $x_0=\frac{\sqrt[3]{5}}{2^{\frac{2}{3}}\sqrt[3]{{{\omega}_{0}}^{2}}}$, then we calculate the Hessian matrix of the potential and evaluate it at $\overrightarrow{x_0}$. The resulting matrix $A$ reads,

\begin{equation}
A={\omega_0}^2
\begin{bmatrix}
14/5 & -8/5 & -1/5\\
-8/5 & 21/5 & -8/5\\
-1/5 & -8/5 & 14/5.
\end{bmatrix}
\end{equation}
Diagonalizing matrix A, we get the normal modes
\begin{equation} 
\Omega_0=\omega_0;\quad \Omega_1=\sqrt{3}\omega_0;\quad  \Omega_2=\sqrt{\frac{29}{5}}\omega_0=\sqrt{5.8}\omega_0.
\label{eq:modes}
\end{equation}
\noindent
The values obtained for the normal modes in Eqs.~(\ref{eq:modes}) coincide with those obtained by Daniel F. V. James in 1997 \cite{ions}.
The first mode $\Omega_0=\omega_0$ is associated with the center of mass. $\Omega_1$ is associated with Coulombian interaction between particles and is called Thomas-Fermi limit \cite{Henning, abraham} or
Universal Breathing Mode (UBM), $\Omega_1$ is present in all electronic systems and is independent of the number of particles $N$ \cite{Henning, Hashemi, Bauch}. 
$\Omega_2$ on the other hand is specific to the 3-fermion one-dimensional harmonic trap \cite{ions}.

With the normal modes obtained in Eqs.~(\ref{eq:modes}) we deduce the spectrum for 3 fermions,

\begin{equation} 
\label{eq:intmcm}
E_{\eta_0\eta_1\eta_2}=V(\overrightarrow{x_0})+{\omega}_{0}(\eta_0+\frac{1}{2})+\Omega_1(\eta_1+\frac{1}{2})+\Omega_2(\eta_2+\frac{1}{2}),
\end{equation}
\noindent

The eigenenergies of the total system $E_{\eta_0\eta_1\eta_2}$ can be expressed as the sum of the eigenenergies of the CM coordinates ${\epsilon_{\eta_0}}^{CM}$ plus the eigenenergies associated with the internal coordinates $q_1,q_2$, namely  $\epsilon_{\eta_1\eta_2}$:
\begin{equation} \label{eigen}
E_{\eta_0\eta_1\eta_2}=\epsilon_{\eta_1\eta_2} + {\epsilon_{\eta_0}}^{CM}=\epsilon_{\eta_1\eta_2} + {\omega}_{0}(\eta_0+\frac{1}{2}), 
\end{equation}
\noindent
with $\eta_0\in\mathbb N$ being the quantum number associated with the CM, and $\eta_1,\eta_2\in\mathbb N$ being the quantum numbers associated to the internal degrees of freedom. Applying Balzer's method only to the internal potential $V(q_1,q_2)$, we obtain a  minimum position $\overrightarrow{q_0}=(q_1=\pm q_0,q_2=0)$, with $q_0=\sqrt{2}x_0$ and

\begin{equation} 
\label{eq:intmcm2}
\epsilon_{\eta_1\eta_2}=V(\overrightarrow{q_0})+\Omega_1(\eta_1+\frac{1}{2})+\Omega_2(\eta_2+\frac{1}{2}),
\end{equation}
\noindent
where $V(\overrightarrow{q_0})$ is the internal potential evaluated at the minimum position $\overrightarrow{q_0}$,
and $\eta_1$,$\eta_2$ represent transitions between internal states.
\begin{table}[]
  \centering
\caption{Analytical values of $\Omega_1$ and $\Omega_2$ for various values of $\omega_0$ (in atomic units)}
  \begin{tabular}{|c|c|c|c|c|}
  \hline
 $\omega_0$ & 0.5 & 0.1 & 0.05 & 0.005  \\ \hline
  $\Omega_1$  & 0.8660 & 0.1732 & 0.0866 & 0.0087 \\ \hline
   $\Omega_2$ & 1.2042 & 0.2408 & 0.1204  & 0.0120 \\ \hline
    
\end{tabular}
\label{Tab:norm_values}
\end{table}

\subsection{Spatial wavefunctions}

According to Balzer \cite{Balzer}, in the strongly correlated limit we can construct 
the total spatial wave function of the system as a multiplication of the elementary wave functions of the harmonic oscillator without Coulombian interaction, decoupled, with each eigenfunction characterized by one of the normal modes $\Omega_i$ in Eqs.~(\ref{eq:modes}),
\begin{equation}
\label{eq:wfBalzer}
\begin{split}
\phi_{\eta_i}(R_i)=\prod_{i=0}^{Nd-1}(\frac{\Omega_i}{\pi\omega_0})^{\frac{1}{4}}\frac{1}{\sqrt{2^{\eta_i}\eta_i!}}e^{\frac{-\Omega_i {R_i}^{2}}{2}}
H_{\eta_i}(\sqrt{\Omega_i}R_i)
\end{split}
\end{equation}
\noindent
with $H_{\eta_i}(\sqrt{\Omega_i}R_i)$ an Hermite polynomial, $N$ the number of particles, $d$ the number of dimensions, and $\overrightarrow{\eta}=(\eta_0,...,\eta_{N-1})$ a set
of labels that represent the transition frequencies
of the total system. In our case we have 3 fermions in one dimension, therefore 3 characteristic frequencies, whose total spatial wave function can be written as the product of
three elementary wave functions of the decoupled harmonic oscillator and $\overrightarrow{\eta}=(\eta_0,\eta_1,\eta_2)$. Notice that the wavefunction Eq.~(\ref{eq:wfBalzer}) is not antisymmetrized, but since in the strongly-correlated limit symmetric and antisymmetric wavefunctions are degenerate in energy the lack of antisymmetrization does not affect the estimation of the energy (see Table \ref{Tab:pot}).  In Table \ref{Tab:hermite} we show the analytic spatial wavefunctions, Eq.~(\ref{eq:wfBalzer}), derived for the lowest 5 spatial wavefunctions of our 3-fermion one-dimensional system, along with the corresponding transition frequencies.


\begin{table*}[t]
\caption{Total spatial wavefunctions in the strongly-correlated limit for 3-fermions trapped in a one-dimensional harmonic trap, derived using Balzer's method \cite{Balzer},  Eq.~\ref{eq:wfBalzer}. The corresponding transition frequencies $(\epsilon_{\eta_0,\eta_1,\eta_2} -\epsilon_{0,0,0})$ are shown in the second column.}
\begin{tabular}{|c|c|}
\hline
Eigenfunctions $\ket{\phi_{\eta_0\eta_1\eta_2}}$ (only spatial part) & frequency   \\ \hline
    $\ket{\phi_{000}}=(\frac{\sqrt{3}\sqrt{5.8}}{\pi^3})^{\frac{1}{4}}{2}^{-\frac{3}{2}}e^{-\frac {(\omega_0{X}^{2}+\Omega_1(q_1-\frac{2x_0}{\sqrt{2}})^{2}+\Omega_2{q_2}^{2})}{2}}$ & 0  \\ \hline
     $\ket{\phi_{100}}=(\frac{\sqrt{3}\sqrt{5.8}}{\pi^3})^{\frac{1}{4}}{2}^{-\frac{1}{2}}\sqrt{\omega_0}Xe^{-\frac {(\omega_0{X}^{2}+\Omega_1(q_1-\frac{2x_0}{\sqrt{2}})^{2}+\Omega_2{q_2}^{2})}{2}}\propto X\ket{\phi_{000}}$ & $\omega_0$ \\ \hline
      $\ket{\phi_{010}}=(\frac{\sqrt{3}\sqrt{5.8}}{\pi^3})^{\frac{1}{4}}{2}^{-\frac{1}{2}}\times{3}^{\frac{1}{4}}\sqrt{\omega_0}(q_1-\frac{2x_0}{\sqrt{2}})e^{-\frac {(\omega_0{X}^{2}+\Omega_1(q_1-\frac{2x_0}{\sqrt{2}})^{2}+\Omega_2{q_2}^{2})}{2}}\propto (q_1-\frac{2x_0}{\sqrt{2}})\ket{\phi_{000}}$ & $\Omega_1$   \\ \hline
       $\ket{\phi_{200}}=(\frac{\sqrt{3}\sqrt{5.8}}{\pi^3})^{\frac{1}{4}}{2}^{-\frac{5}{2}}(4\omega_0X^{2}-2)e^{-\frac {(\omega_0{X}^{2}+\Omega_1(q_1-\frac{2x_0}{\sqrt{2}})^{2}+\Omega_2{q_2}^{2})}{2}}\propto (4\omega_0X^{2}-2)\ket{\phi_{000}}$ & $2\omega_0$\\ \hline
        $\ket{\phi_{001}}=(\frac{\sqrt{3}\sqrt{5.8}}{\pi^3})^{\frac{1}{4}}{2}^{-\frac{1}{2}}\times{5.8}^{\frac{1}{4}}\sqrt{\omega_0}q_2e^{-\frac {(\omega_0{X}^{2}+\Omega_1(q_1-\frac{2x_0}{\sqrt{2}})^{2}+\Omega_2{q_2}^{2})}{2}}\propto q_2 \ket{\phi_{000}}$ & $\Omega_2$  \\ \hline
\end{tabular}
\label{Tab:hermite}
\end{table*}

\subsection{Ground state density}

As $\omega_0\to0$ and repulsion dominates over confinement, the fermions localize maximally far from each other. This regime is known as Wigner crystal in the condensed matter community \cite{Cioslowski, plata, wigner, wigner_2}.  There has been recent experimental evidence of this phenomenon \cite{wigner_exp}.

To compute the ground state density we adapted Balzer's Equation (24) in Ref.~\cite{Balzer} to 3 fermions:

\begin{equation}\label{eq:densb} 
\rho_{\overrightarrow{\eta}}(x)
=\int{dx_1dx_2dx_3|\Phi_{\overrightarrow{\eta}}(x_1,x_2,x_3)}|^{2}
\sum_{i=1}^{3}\delta(x_i-x) .
\end{equation}
\noindent
After replacing the wave functions with Eq.~(\ref{eq:wfBalzer}) we obtain the density as a sum of 3 Gaussian functions:
\begin{equation} 
\label{eq:dens} 
\begin{split}
\rho(x)=A_1(\omega_0,\Omega_1,\Omega_2)e^{-f_1(\omega_0,\Omega_1,\Omega_2)(x-x_0)^{2}}
\\+A_2(\omega_0,\Omega_2)e^{-f_2(\omega_0,\Omega_2)x^{2}}
\\+A_1(\omega_0,\Omega_1,\Omega_2)e^{-f_1(\omega_0,\Omega_1,\Omega_2)(x+x_0)^{2}},
\end{split}
\end{equation}
\noindent
with $f_1(\omega_0,\Omega_1,\Omega_2)=\frac{\Omega_1\Omega_2}{\frac{\Omega_2}{2}+\frac{\Omega_1}{6}+\frac{\Omega_1\Omega_2}{3\omega_0}}$ and $f_2(\omega_0,\Omega_2)=\frac{3\Omega_2}{2+\frac{\Omega_2}{\omega_0}}$.According to the definition of a Gaussian function: $\sigma_i^{2}=\frac{1}{2f_i(\omega_0,\Omega_1,\Omega_2)}$, and $A_i(\omega_0,\Omega_1,\Omega_2)=\frac{1}{\sigma_i\sqrt{2\pi}}$.
The first and third terms represent each of the end bell curves, and
the second term represents the middle bell curve. It can be noted that the number of peaks appearing in the density is equal to the number of fermions in the system. 
\begin{figure}[htbp]
    \centering
    \includegraphics[width=0.48\textwidth]{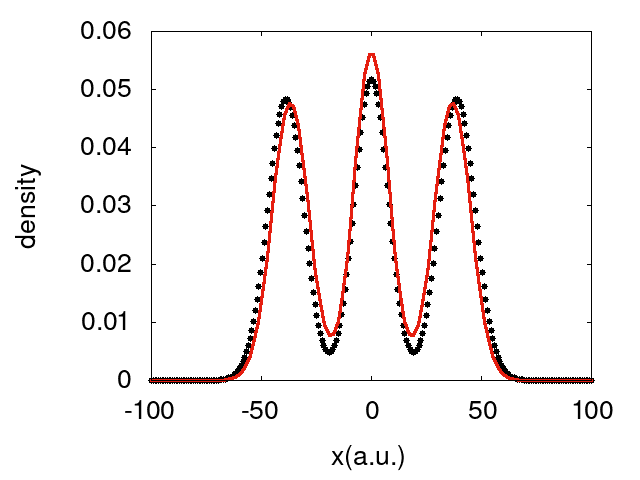}
    \caption{Comparison between the ground state density of the 3-fermion one-dimensional harmonic trap with $\omega_0 = 0.005$ in the purely symmetric state, obtained by numerical simulation in octopus (black dotted), and the analytic expression found in the limit of $\omega_0\to 0$ Eq.~(\ref{eq:dens}) (red line).}
    \label{fig:balz_num}
    \end{figure}

\begin{figure}[htbp]
    \centering
    \includegraphics[width=0.4\textwidth]{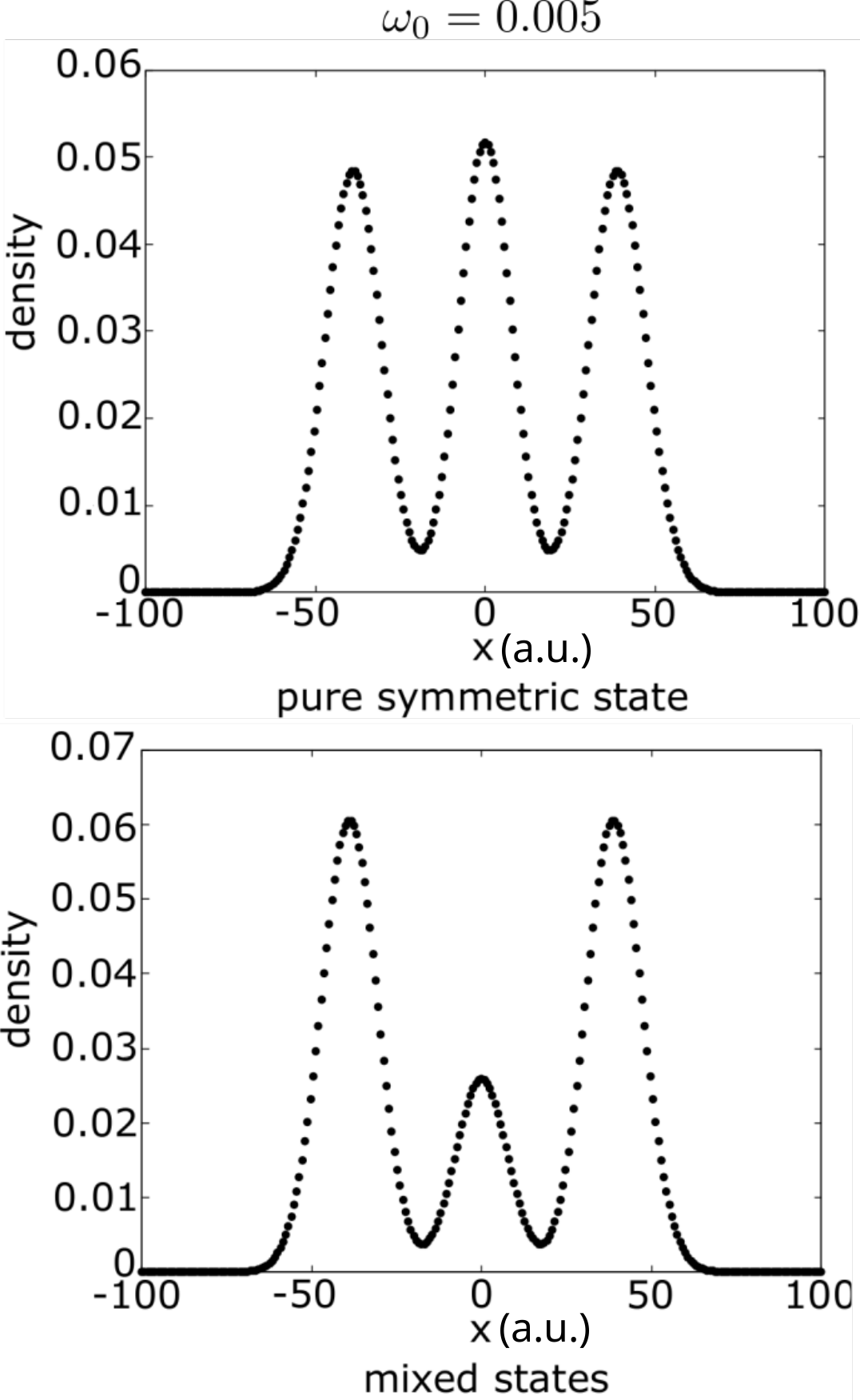}
    \caption{Ground state density of the 3-fermion one-dimensional harmonic trap with $\omega_0 = 0.005$ in the purely symmetric (upper panel) and mixed (lower panel) cases of lowest energy. The calculations were performed in Octopus using a simulation box of size $acell = 100$ a.u and $spacing = 0.8$ a.u.}
    \label{fig:densall}
    \end{figure}

    \begin{figure}[htbp]
    \centering
    \includegraphics[width=0.48\textwidth]{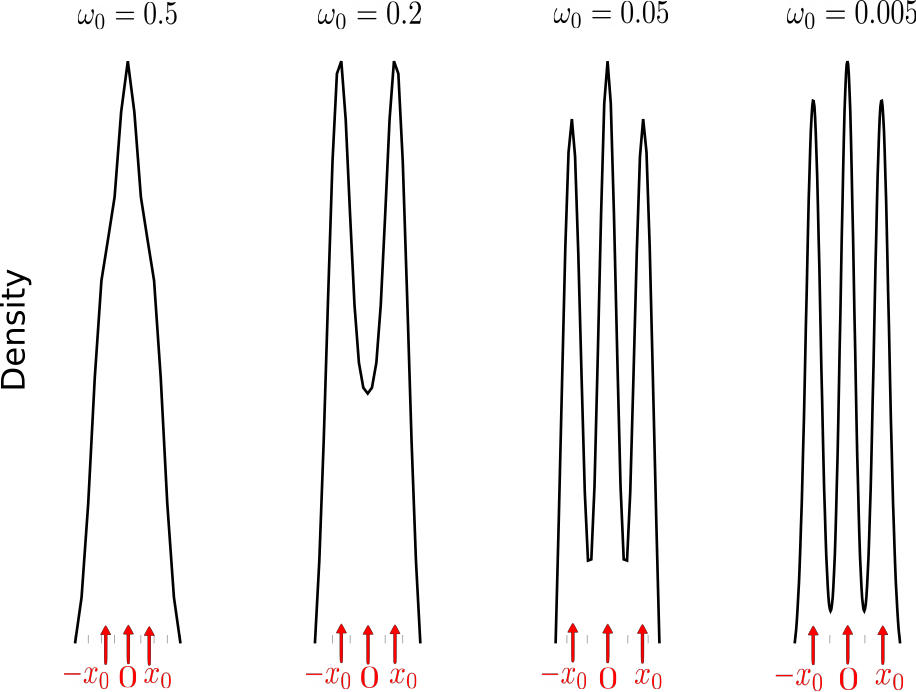}
    \caption{The numerically simulated ground state density for various correlation regimes corresponding to increasingly weaker trapping potentials (increasingly smaller $\omega_0$), purely symmetrical case. }
\label{fig:dens}
    \end{figure}
\begin{figure*}[htbp]
    \centering
    \includegraphics[width=0.75\textwidth]{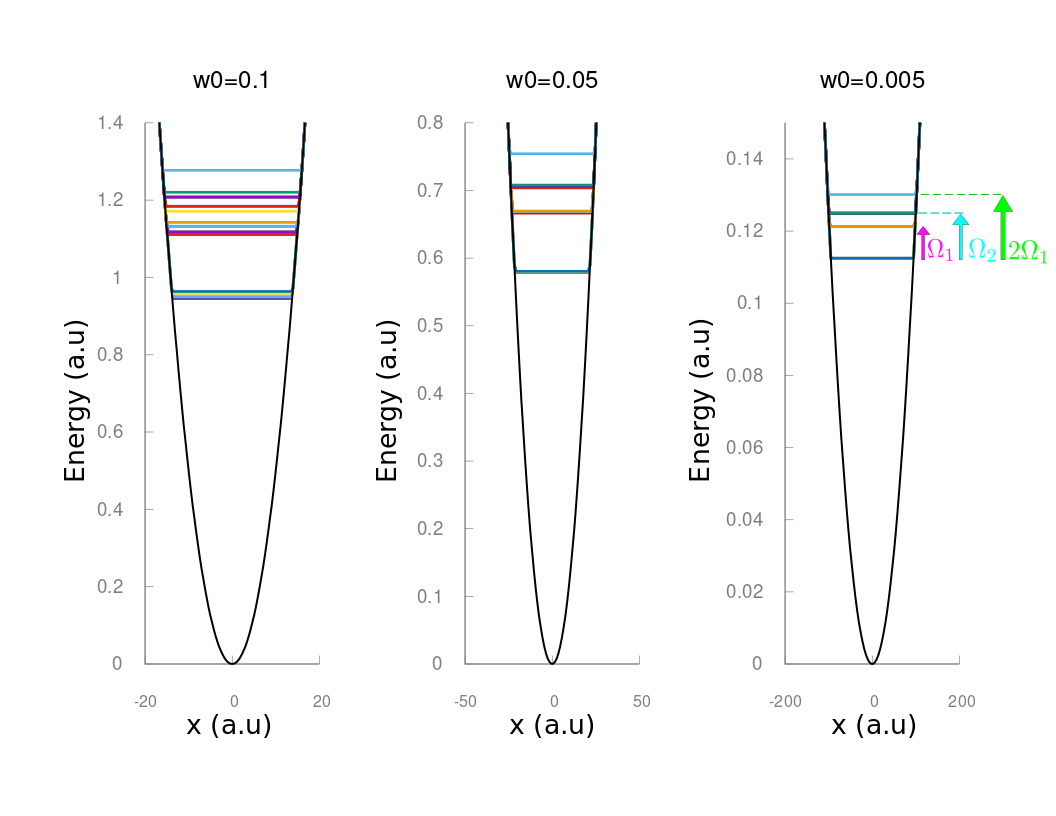}
    \caption{ The internal spectrum of the system for 3 different values of $\omega_0$ to observe the transition towards the high correlation limit. At $\omega_0=0.005$ it is observed
that the spacing between eigenenergies is equal to the frequencies of the normal modes.}
    \label{fig:esp}
\end{figure*}

\section{Numerical simulations}
\label{sec:numerical}

The total Hamiltonian as well as the CM and internal Hamiltonians, $H_{CM}(X)$, $H_q(q_1,q_2)$ and $H_{Total}(X,q_1,q_2)=H_{CM}(X)+ H_q(q_1,q_2)$ (Eqs.~\ref{eq:Hcm}-\ref{eq:Hq}), were solved numerically using the open-source software Octopus \cite{Octopus}, with the interaction between particles modeled as Soft-Coulomb,

\begin{equation}\label{vint}
\begin{split}
    V_{S-Cou}(q_1,q_2)=\frac{1}{\sqrt{(\sqrt{2}q_1)^{2}+{a}^{2}}} 
    \\ + \frac{1}{\sqrt{(\frac{-q_1+\sqrt{3}q_2}{\sqrt{2}})^{2}
   +{a}^{2}}}
    \\+ \frac{1}{\sqrt{(\frac{q_1+\sqrt{3}q_2}{\sqrt{2}})^{2}+{a}^{2}}},
    \end{split}
\end{equation}
to avoid the singularities in the one-dimensional Coulombian potential \footnote{The Soft-Coulomb parameter $a$ models interaction
properties. In the case of quantum wires, the parameter $a$ mimics the short-range cutoff distance caused by the transverse dimension of the system (for example the diameter of a carbon nanotube or a quantum wire semiconductor) \cite{plata}} \cite{plata, soft_2}. The parameter $a$ is the Soft-Coulomb parameter, which we set to $a=1$. 

In Table \ref{Tab:pot} we compare the internal eigenenergies derived in the strong correlation limit for Coulomb and Soft-Coulomb interaction with the ones obtained numerically with Soft-Coulomb and a finite but small $\omega_0$ in Octopus. The analytic Coulomb versus Soft-Coulomb eigenergies differ by less than $10^{-4}$ Hartree. These results confirm the overlap between the three wavefunctions is very small (Wigner Crystal). Because of the small overlap the lack of antisymmetrization of Eq.~(\ref{eq:wfBalzer}) does not affect the ground state energy. The results also show the approximation introduced by the Soft-Coulomb potential is very small.
The discrepancy between the analytic and numerical soft-Coulomb solutions, of the order of $10^{-3}$ Hartree, is due to the finite value of $\omega_0$ used in the simulations, as can be seen from the fact that the discrepancy is smaller for $\omega_0=0.001$ than it is for the less correlated $\omega_0=0.005$ system. 
\begin{table}[]
\centering
\scriptsize
\caption{Internal eigenenergies derived for the strongly-correlated limit for Coulomb interaction (4th column) and Soft-Coulomb interaction (5th column) versus numerical solution for $\omega_0=0.005$ and $\omega_0=0.001$ and Soft-Coulomb interaction  (6th column)}
\begin{tabular}{|c|c|c|c|c|c|}
\hline
Corr. regime & $\eta_1$ & $\eta_2$ & $\epsilon_{\eta_1\eta_2}$  & $\epsilon_{\eta_1\eta_2}$ S-Cou & Numerical\\ \hline
  $\omega_0=0.005$ & 0  & 0 & 0.112142 & 0.112120 & 0.112465\\ \hline
    & 1  & 0 & 0.120802 & 0.120781 & 0.121274\\ \hline
    & 0  & 1 & 0.124184 & 0.124162 & 0.124927 \\ \hline
     & 2  & 0 & 0.129463 & 0.129441 & 0.130133 \\ \hline
      & 1 & 1 & 0.132844 & 0.132822 & 0.133879 \\ \hline
       & 0  & 2 & 0.136225 & 0.136204 & 0.137636 \\ \hline
       \hline
        $\omega_0=0.001$ & 0  & 0 & 0.034561 & 0.036881 & 0.036921\\ \hline
    & 1  & 0 & 0.036293 & 0.038613 & 0.038671\\ \hline
     & 0  & 1 & 0.036969 & 0.039290 & 0.039380 \\ \hline
     & 2  & 0 & 0.038025 & 0.040345 & 0.040427 \\ \hline
       
\end{tabular}

\label{Tab:pot}
\end{table}
Since the overlap between the three wavefunctions is very small (Wigner Crystal), the lack of antisymmetrization of Eq.~(\ref{eq:wfBalzer}) does not affect the ground state energy, which is well captured by this model, as can be confirmed by the similar results obtained for analytic and numerical solutions in Table \ref{Tab:pot}.

\subsection{Ground state density}
In Fig.~\ref{fig:balz_num} we compare the density derived for the strong correlation limit Eq.~(\ref{eq:dens}) with a numerically simulated density for a small $\omega_0$. We find two distinct solutions for the numerical ground state density, corresponding to the purely symmetric and mixed fermionic wavefunctions, depicted for the strongly-correlated $\omega_0=0.005$ a.u. case in Fig.~\ref{fig:densall}. The relationship between the heights of the three peaks in the density follows the pattern described in \cite{plata,oro}. In the strongly correlated regime, the pure symmetric states shows a subtle difference in the peak heights, with all three peaks being of similar height (Fig.1 in \cite{oro}). For mixed states, peaks with more differentiated heights are observed in the density, whith the middle peak being much shorter than the external peaks.
 We notice that the analytic wavefunctions Eq.~(\ref{eq:wfBalzer}) are not antisymmetrized, therefore there is a unique solution instead of two. Comparing the analytic and numerical densities we observe that the Balzer wavefunctions best represent the pure antisymmetric case.
For the aim of comparison, we present the equilibrium positions $\overrightarrow{x_0}$ and the width  $\sigma$  of the bell curves calculated from Eq.~(\ref{eq:dens}) in Tables \ref{Tab:xov} and \ref{Tab:sig} respectively.
In Fig.~\ref{fig:dens} we show the purely symmetrical ground state density for different correlation regimes corresponding to increasing correlated systems $\omega_0=0.5;0.2;0.05;0.005$ a.u. We observe that as we approach the strongly correlation limit the number of peaks in the density grows from one to two to three.

 \begin{table}[]
\caption{$\omega_0=0.005$ a.u. (strong correlation regime): Absolute value $|x_0|$ (a.u.) of the equilibrium position  $\protect\overrightarrow{x_0}=(-x_0,0,x_0)$, from the generalization of Balzer's solution to 3 fermions (analytic, not anisymetrized) versus numerical solution for symmetric and mixed cases.}
  \begin{tabular}{|c|c|c|c|}
  \hline
$\omega_0$ & $\overrightarrow{x_0}$ analytic& $\overrightarrow{x_0}$ symmetric & $\overrightarrow{x_0}$ mixed\\ \hline
 0.05 & 7.937 & 8.739& 8.739 \\ \hline
     0.005 & 36.840 & 38.907 & 38.926 \\ \hline
\end{tabular}
\label{Tab:xov}
\end{table}

\begin{table}[]
  \centering
\caption{Width Comparison ($\sigma$) of the density bell curves obtained analytically and numerically for a value of $\omega_0=0.005$ (high correlation).}
  \begin{tabular}{|c|c|c|c|c|}
  \hline
 $\omega_0$ & Bell curves & $\sigma_{analitico}$ & $\sigma_{sim}$ & $\sigma_{mix}$ \\ \hline
   0.05 & Middle & 2.470  & 2.099 & 2.099 \\ \hline
   0.05 & End & 2.629  & 2.707 & 2.707 \\ \hline
   0.005 & Middle & 7.8113  & 7.4129 & 6.9870 \\ \hline
    0.005 & End &  8.2705&  8.2676 & 8.2956\\ \hline
\end{tabular}
\label{Tab:sig}
\end{table}

\subsection{von Neumann entropy}

In the case of the 2-fermion trap it was observed that, as we approach the strongly-correlated regime, the von Neumann entropy grows \cite{ic1/ic2}.
We found that this is also the case for the 3-fermion trap and is likely to be the case for any $N$. 
The von Neumann entropy is a measure of the entanglement present in the system \cite{entropy}. In this problem, strong-correlation implies large entanglement, two deeply related concepts that are sometimes elusive to differentiate. 
In Table \ref{Tab:neu} we show the von Neuman entropy $s^{\omega_0}$ for various values of $\omega_0$, 
\begin{equation}
\label{entropy}
    s^{\omega_0}=\frac{1}{N}\sum_{i=1}^N \eta^{\omega_0}_i ln(\eta^{\omega_0}_i)
\end{equation}
where $\eta^{\omega_0}_i$ is the Natural Orbital Occupation Number of Natural Orbital $i$ , and $N$ is the number of particles in the system. To obtain $\eta^{\omega_0}_i$ we diagonalized the 1-Body Reduced Density Matrix $\Gamma^{(1)}(x;x')$ for various $\omega_0$,

\begin{equation}
\begin{split}
\Gamma_{\omega_0}^{(1)}(x;x')&=\int dx_2 dx_3  \Psi_{\omega_0}^*(x_1', x_2',x_3') \Psi_{\omega_0}(x_1, x_2,x_3)
\\
&=\sum_{j=1}^\infty \eta_j^{\omega_0}\varphi_j^{\omega_0}(x)\varphi_j^{\omega_0,*}(x') 
\end{split}
\end{equation}


\begin{table}[]
  \centering
\caption{Von Neumann entropy for various values of $\omega_0$ in a.u.}
  \begin{tabular}{|c|c|c|c|c|}
  \hline
 $\omega_0$ & 0.5 &  0.1 & 0.05 & 0.005  \\ \hline
   $s^{\omega_0}$ & 0.169 & 0.329 & 0.367 & 0.386  \\ \hline
    
\end{tabular}
\label{Tab:neu}
\end{table}
Notice that in the Hartree-Fock and DFT methods the von Neumann entropy is always zero, because all $\eta_i$ are either 0 or 1 \cite{Natorbs}.  Thus for these efficient ab-initio methods simulating the strongly-correlated regime is a challenge. In Ref.~\cite{ic1/ic2} it was shown that DFT using LDA approximation fails to capture the localization of the density in the strongly correlated limit in the 2-fermion harmonic one-dimensional harmonic trap case (see inset Fig.3 in Ref.~\cite{ic1/ic2}). Even if other exchange-correlation approximations may capture the strong-correlation limit \cite{Malet12} the simulation of the transition from a weakly to a strongly-correlated regime (for example dissociation curves of molecules) remains a challenge.  

\subsection{Energy spectrum and selection rules}
The internal energy spectrum (ground state plus 18 unoccupied states) was calculated numerically in Octopus and is shown in Fig.~\ref{fig:esp}.  We observe that as we make $\omega_0$ smaller and approach the strongly-correlated regime, the spacing between internal eigenenergies converges to the normal modes.
for computational efficiency we calculated the spectrum of the internal system instead of the spectrum of the total system. Since the Hamiltonian is separable the total response of the system is the sum of the CM and internal responses. The CM spectrum is that of a harmonic oscillator of natural frequency $\omega_0$.

To measure the internal response we performed a time evolution of the internal density after perturbation with an instantaneous quadrupole external field (quadrupole kick).   
For a sufficiently small applied field, the response can be expanded in a series of powers in the intensity $E_0$ of the field. The linear ($\propto E_0$)  
response of an observable $\hat{O}$ is proportional to the matrix element $\braket{{\hat{O}}^{(1)}}\propto \sum_{k=1}^{N}\bra{\phi_0}\hat{O}\ket{\phi_k}\bra{\phi_k}\hat{H_I}\ket{\phi_0}$ 
\cite{ic1/ic2}\cite{hyp}.
By performing the Fourier transform of the evolved density, we obtain the frequency spectrum, where the peaks correspond to transition frequencies between the ground and excited states that are not forbidden by symmetry. The allowed transitions depend on the symmetry of the system and also of the excitation field (selection rules). Notice that to excite internal degrees of freedom and study correlation effects we need a quadrupole or higher field because a dipole field (plane waves) can only excite CM transitions (Kohn's theorem \cite{Yip, Brey}).
 To study correlation effects in the quadrupole spectrum, a light field with spatial structure, for example, optical vortices (also known as twisted light), is proposed, the details of the setup are discussed in the next section. 

\begin{figure}[htbp]

   \centering
    \includegraphics[width=0.48\textwidth]{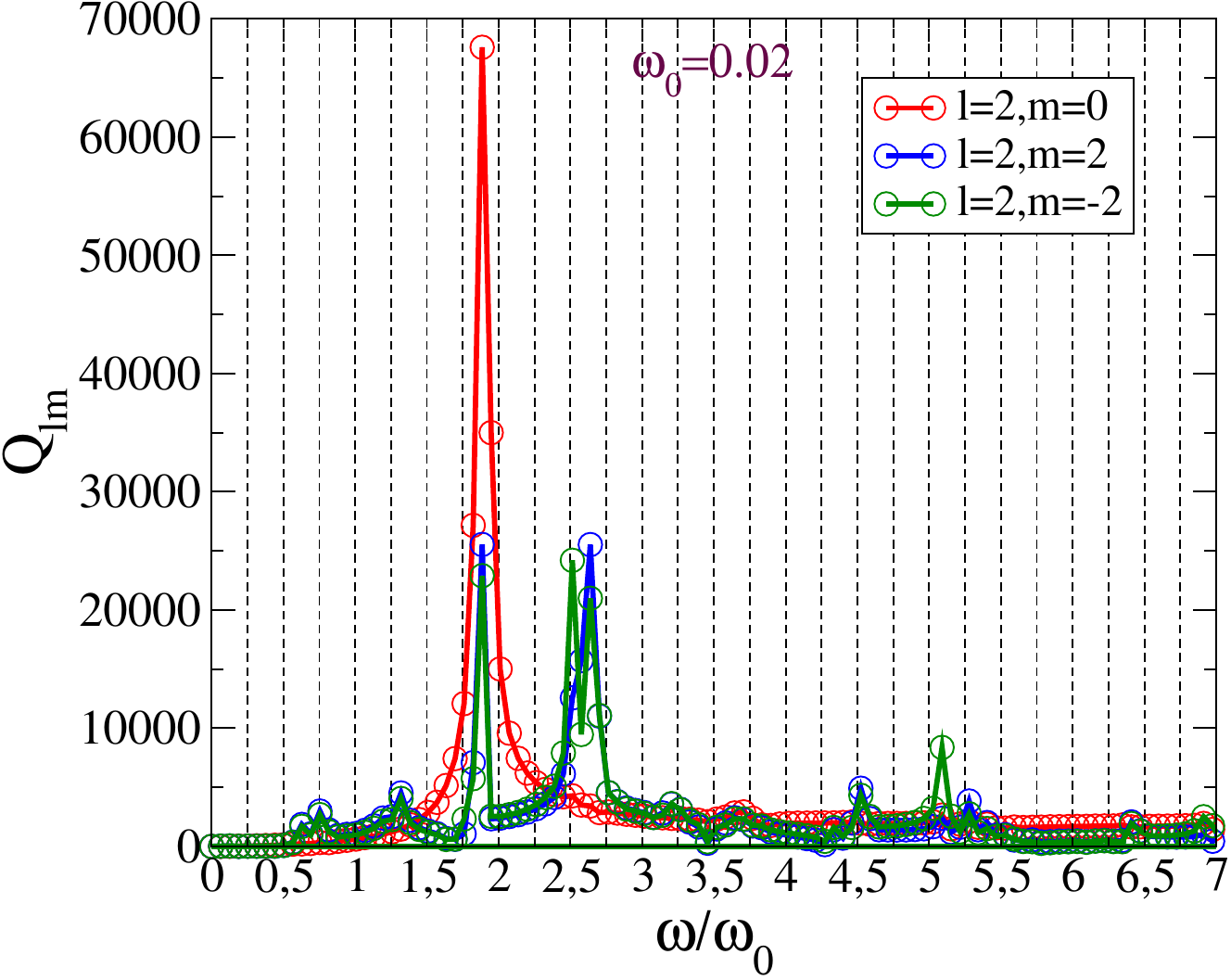}
   \includegraphics[width=0.48\textwidth]{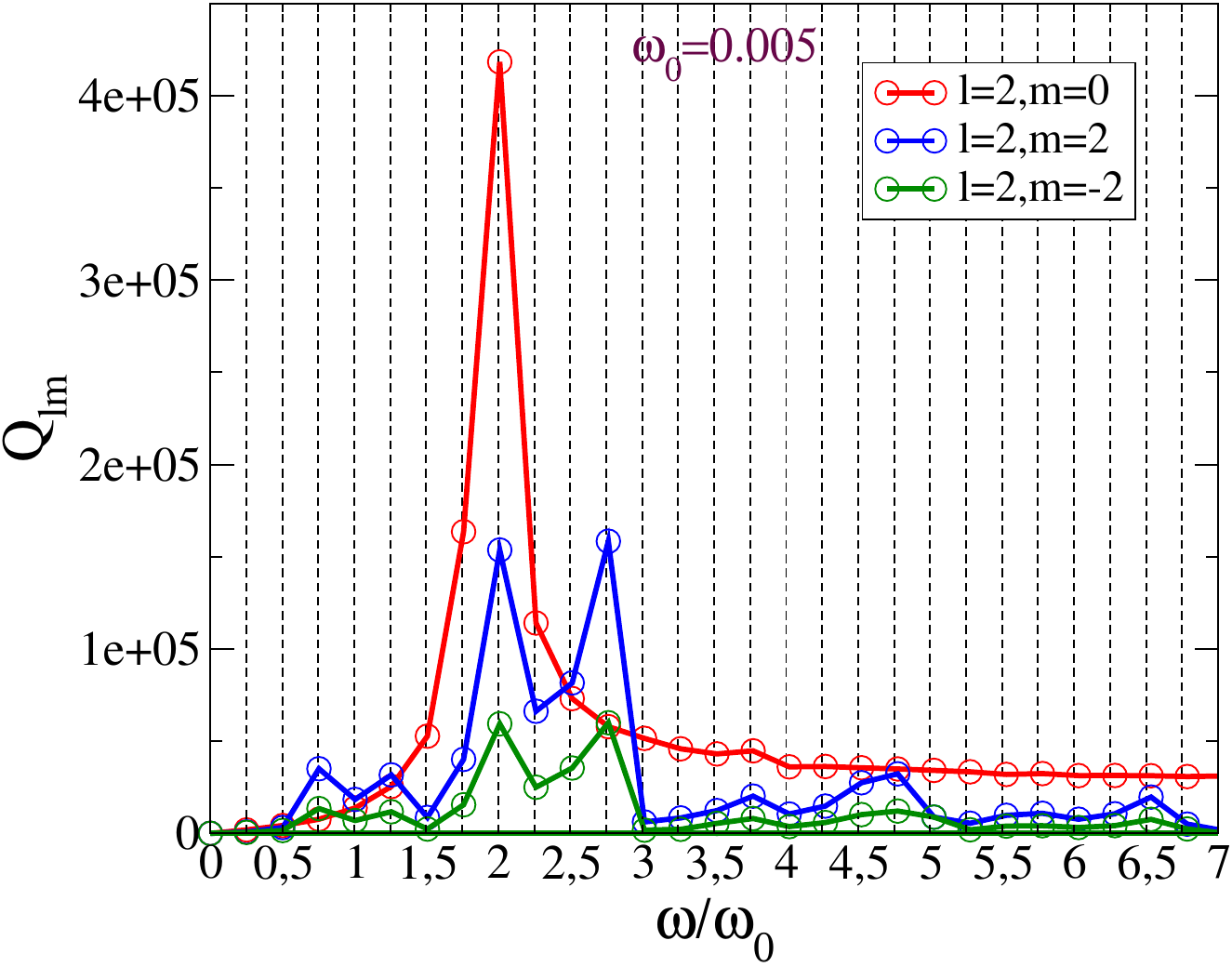}
   \includegraphics[width=0.48\textwidth]{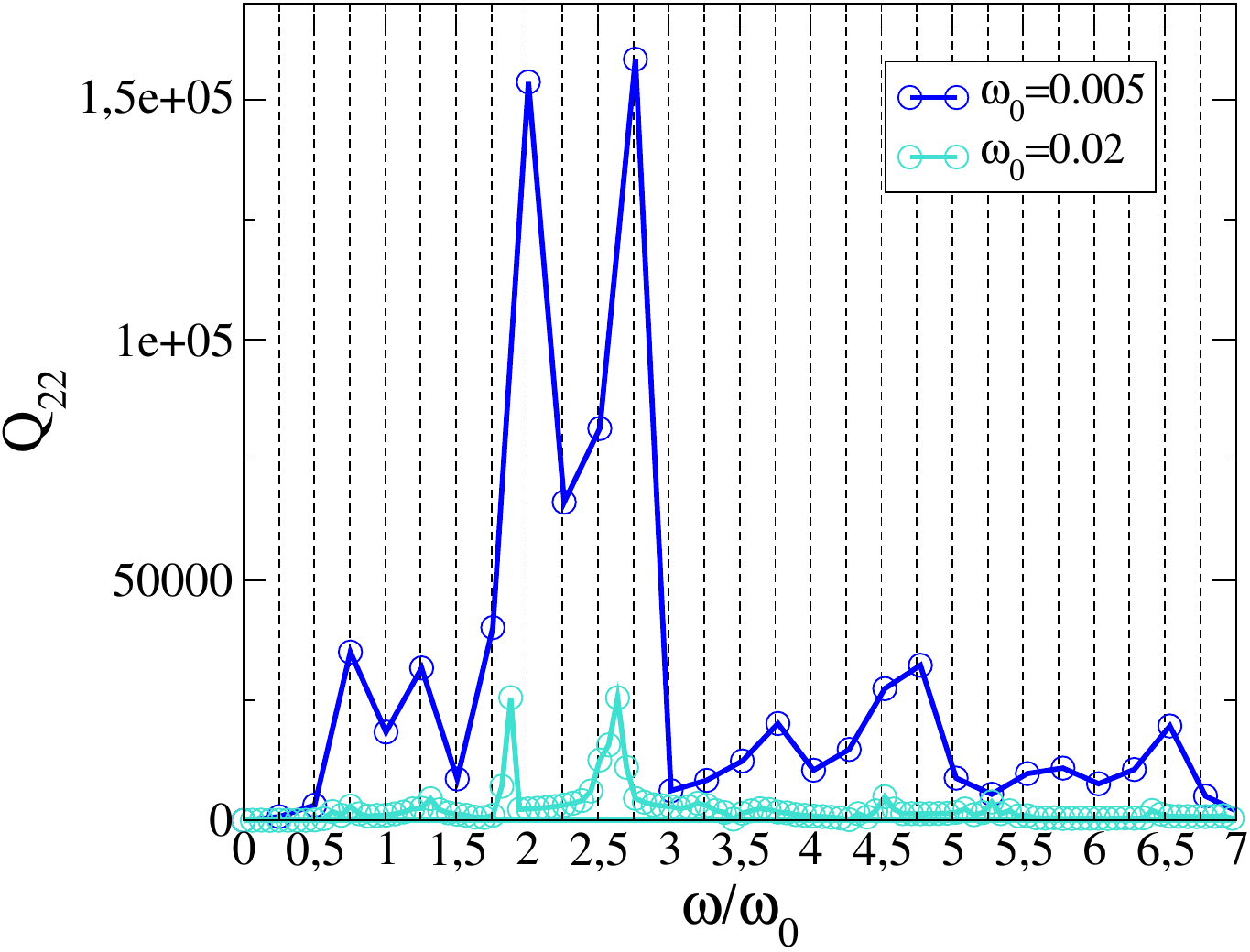}
  
   \caption{Quadrupole spectra (in arbitrary units). The frequency is divided by $\omega_0$. Upper panel $\omega_0=0.02$, middle panel $\omega_0=0.005$ a.u. for various values of the spin angular momentum $m$: $Q_{m=0}$ (red), $Q_{m=2}$ (blue) and $Q_{m=-2}$ (purple). Lower panel: comparison of $Q(l=m=2)$ for $\omega_0=0.005$ (blue) versus $\omega_0=0.02$ a.u. (cyan).} 
    \label{fig:esp_2}
\end{figure}

In Fig.~\ref{fig:esp_2} we show the internal quadrupole spectrum obtained from Fourier transformation of the time-resolved internal density for $\omega_0=0.02$ and $\omega_0=0.005$ a.u.. The computational details are included in the Appendix.
We plot the absolute value of the quadrupole spectrum against a dimensionless frequency (expressed in units of  $\omega_0$).
The peaks are close but not on top for the two values of $\omega_0$. We notice that for $\omega_0=0.02$ a.u. there is an additional peak, this is because for this less correlated system, the spectrum is not yet the simple equidistant spectrum of the strongly-correlated limit (see Fig.~\ref{fig:esp}).  
 We observe that the position of the peaks (corresponding to the allowed transition frequencies) approximately obey: $n.(\Omega_1+\Omega_2)/(3!\omega_0)= n.(\sqrt{3}+\sqrt{5.8})/6= n.\Omega_{eff}$, with $n\in\mathbb Z$ and $3!=N!$. This would mean that there is only one independent internal quantum number $n_{eff}$ instead of two.
 In Table \ref{Tab:peaks} we show the numerical values of the peak positions obtained by Octopus and compare them with the analytical values. We observe that
for $n_{eff}=3;4$ the behavior of $n_{eff}$ is reassuring, namely the more correlated $\omega_0=0.005$ system shows very good agreement with the strong-correlated limit analytic solution, better than the less correlated $\omega_0=0.02$ system.  But the behavior of $n_{eff}=1;2$ doesn't fit so well into the hypothesis of an equidistant spectrum.
The explanation of these observations is a work in progress and will be the topic of a future publication.

\begin{table}[]
  \centering
\caption{Peak positions for $\omega_0=0.02$ and $\omega_0=0.005$ a.u. (in units of $\omega_0$). The first column is the internal quantum number $n$, such that the peaks are at $n \Omega_{eff}$ (see discussion in the text).}
  \begin{tabular}{|c|c|c|c|c|}
  \hline
 n & $\omega_0$  & Numerical & Analytic ($\omega_0\to 0)$) \\ \hline
  1 & 0.02  &  0.6284 & 0.6901 \\ 
   &  0.005   & 0.7540 & 0.6901 \\ \hline
 2 &  0.02  &  1.3198 & 1.3801 \\ 
   & 0.005 &    1.2567 & 1.3801 \\ \hline
  3 & 0.02  &  1.8845 & 2.0702 \\ 
   & 0.005   &  2.0101 & 2.0702 \\ \hline
   4 & 0.02 &  2.6389 & 2.7602 \\ 
    & 0.005   &  2.7647 & 2.7602 \\ \hline
\end{tabular}
\label{Tab:peaks}
\end{table}


\section{Measurement of the quantum observables}
\label{sec:exp}

The transition into a strongly-correlated regime can be observed in the absorption spectrum, i.e. the response of the system to a perturbation with a light field.
A homogenous light field, however, can only excite the CM degrees of freedom in harmonic traps, because it couples to the dipole moment of the system, which is a CM variable (see Kohn's theorem \cite{Yip}). We therefore need a quadrupole field or higher to study the correlation between particles, such that internal degrees of freedom do get excited.
We propose to use the spatial structure of a twisted light field to excite dipole-forbidden transitions and study the correlation effects \cite{ic1/ic2,qui_3,forb,forb_2,ov,ov_2}. 
Such a setup could be used to probe the correlation regime but also to selectively excite the system into a desired dipole-forbidden state, i.e. preparing the system in a particular initial state. The latter may have applications in digital ion trap quantum computers, whose working regime is the strongly correlated limit we have investigated in this paper \cite{Wineland98,IonSimu}
 
Twisted light, also known as optical vortices, 
designates a family of highly non-homogeneous optical beams with single or multiple phase singularities carrying orbital angular momentum (OAM), among other interesting features \cite{ov,ov_2,allen03,andrews11,bliokh15, qui, qui_2}. 
We propose to use the spatial inhomogeneity associated with the topological charge $l$ of the optical vortices to study the internal energy spectrum of the trap. This idea was first introduced in Ref.~\cite{ic1/ic2} to study 2 harmonically trapped fermions. We model the interaction between the system and an optical vortex of topological charge $l=1$ and $l=2$ as, 
\begin{equation} \label{hl1} 
{H_I}^{l=1}=\frac{eE_0q_r}{2\sqrt{2}}(X^2+{q_1}^2+{q_2}^2)sen(\omega t)
\end{equation}
\begin{equation} \label{hl2} 
{H_I}^{l=2}=\frac{eE_0{q_r}^2}{12\sqrt{2}}X(X^2+{q_1}^2+{q_2}^2)sen(\omega t)
\end{equation}
where $E_0$ and $q_r$ are the amplitude and the waist of the beam respectively. 
Details on the derivation of Eqs.~(\ref{hl1})(\ref{hl2}) and the experimental setup for the twisted light probe can be found in Ref.~\cite{ic1/ic2}. The specifications for the twisted light probe depend on the size of the system (which can be estimated from the density) rather than on the number of fermions in the trap.

In Fig.~\ref{fig:esp_2} we present the internal quadrupole spectrum of a 3-fermion harmonic trap after perturbation with a twisted light field of $l=1$, whose interaction with the system is described by Eq.~(\ref{hl1}). The CM quadrupole spectrum has peaks at even multiples of the natural frequency, i.e. $2\eta_0 \omega_0$ where $\eta_0$ is the CM quantum number (not shown).


\section{Conclusions} \label{sec:conclusions}

We were able to elucidate 4 features that characterize the strongly-correlated regime in one-dimensional fermion harmonic traps.  Our findings could be used to probe the interactions between the electrons in quantum dots or between the ions in Paul traps. 
 Any of these platforms could be used as an analogue quantum computer to simulate the transition into a strongly correlated regime in more complex quantum many-body systems, something hard to explore with classical processors.
Wigner crystallization \cite{wigner, wigner_2, wigner_exp}, bosonization \cite{Z_on_cold_atoms,boson,boson_2,boson_3,boson_4,Lindsay,boson_5,boson_6}, and metal-insulator transitions \cite{Feng21} are some examples of phenomena that can emerge as we approach the strongly-correlated regime.

In addition, identification of the characteristic signatures of strong correlation allows us to identify the quantum observables we need to track and is helpful for 
the interpretation of experimental results on strongly-correlated materials.
Some of the features we identified as signatures of the strongly-correlated regime 
are common to strongly-correlated materials relevant to the industry, 
such as transition metals (used as catalysts in energy storage devices) and stretched molecules involved in chemical reactions.
In these systems, Hartree Fock is a bad approximation (or what is the same, the von Neumann entropy is non-zero) and the energy levels become degenerate, which makes convergence for ab initio methods such as DFT, challenging  \footnote {Degenerate states make the convergence of the self-consistent field calculation in HF and DFT hard, requiring tricks such as level shift or smearing.}.

In this work we generalize the findings of \cite{ic1/ic2} from 2 to 3-fermion one-dimensional traps. We expect the same signatures of strong correlation to be present in the more general case of $N$-fermion traps. Generalization to higher dimensional traps is the scope of future work.

We have derived an analytical expression for the spatial part of the 3-fermion wavefunction in the highly correlated limit
and we show that, in this limit, the energy spectrum of the 3-fermion one-dimensional harmonic trap is fully determined by 3 normal modes: 
two coincide with the modes of the 2-fermion trap \cite{ic1/ic2,Balzer,Taut_2}, namely $\omega_0$ associated with the center of mass and the universal breathing mode $\Omega_1=\sqrt{3}\omega_0$, and the third normal mode $\Omega_2=\sqrt{5.8}\omega_0$ is specific to the 3-fermion one-dimensional trap \cite{ions}. 
The many-body wavefunction can be approximated by a product of Gaussian functions\cite{Tomasz} (independent oscillators with characteristic frequencies equal to the normal modes) \cite{entang,Balzer}. We used these analytic results to validate the numerical simulations.

The ground state wavefunction was simulated numerically using soft-Coulomb interaction and the open source Octopus code \cite{Octopus} for 
different values of the trap frequency $\omega_0$. As a check, we ensured the simulated internal wavefunction respects parity (see Fig.~\ref{fig:intw} in the Appendix). We also checked that the use of Soft-Coulomb interaction, instead of full Coulomb, does not significantly affect the density. We found good agreement between the analytical density in the strongly-correlated limit and the simulated symmetric ground state density for small but finite  $\omega_0$ (see Fig.\ref{fig:balz_num}).

One of the 4 characteristic features of the strongly-correlated limit is that the fermions localize maximally far from each other (Wigner Crystal, see Fig.~\ref{fig:dens}). This localization renders the fermions distinguishable, because it becomes possible to label them. We show that for 3 fermions the total wavefunction can be written as a product of 3 Gaussian functions with little overlap among them in the Wigner Crystal regime. As a consequence, the antisymmetrization of the wavefunction does not affect the energy, and thus symmetric (bosons) and antisymmetric (fermions) solutions become degenerate, which is the second feature we identify as characteristic of the strongly-correlated limit. There has been experimental evidence of this symmetry \cite{Z_on_cold_atoms, Zuern15}, also known as bosonization \cite{Z_on_cold_atoms,boson,boson_2,boson_3,boson_4,Lindsay,boson_5,boson_6}. 
The third characteristic feature we have identified is the growth of the van Neumann entropy \cite{entropy} (see table \ref{Tab:neu}), which is a measure of the entanglement present in the system.
The forth characteristic feature is the simplification of the energy spectrum. As we approach $\omega_0$ the spectrum is fully characterized by $N$ characteristic frequencies corresponding to the $N$ normal modes.
For intermediate $\omega_0$ the energy spectrum has many more distinct energy levels (see Fig.~\ref{fig:esp}). 

To study the fingerprints of particle-particle correlation in the spectrum we perturbed the system with a quadrupole field. A twisted light probe could be used for this purpose as proposed in Ref.~ \cite{ic1/ic2,quadru}. 
We observe that as we approach $\omega_0\to 0$, the peaks in the simulated spectrum approach combinations of the normal modes $\omega_0$, $\Omega_1$, $\Omega_2$, as expected. 
The positions of the peaks in the simulated quadrupole spectrum (see Fig.~\ref{fig:esp_2}) are observed to be approximately proportional to the sum of $\Omega_1$ and $\Omega_2$ divided by the factorial of the number of particles $N!=3!$, which suggests that we can define this sum as an effective internal natural frequency $\Omega_{eff}=(\Omega_1+\Omega_2)/3!$. The explanation of these selection rules is the scope of future work.

\section{Appendix: Young tableaux of the wavefunctions of the system of 3 electrons.} \label{sec:y}

A Young diagram is a mathematical object made up of rows of boxes, the arrangement of boxes is such that each row has a quantity less than or equal to the top row. This object is related to Group Theory.
A Young Tableau is defined through the placement of symbols, particularly numbers, into the boxes of the corresponding Young diagram\cite{young}.

\ytableausetup{centertableaux}
\begin{ytableau}
1 & 2 & 3 & 4 & 5 \\
6 & 7 & 8 \\
9 & 10  \\
\end{ytableau}

Different diagrams have several standard tableaux, which are the result of the different ways of arranging the numbers in Young's tableaux \cite{Ma,young} but arranged in such a way that an ascending sequence can be read both by rows (from left to right) and columns (from top to bottom). For example, for the previous example, there are 10 standard tableaux.

For 2 electrons, two diagrams can be associated with the singlet and triplet configurations, respectively.
When there are 3 electrons involved, there are two diagrams in which 2 electrons occupy one spin channel while the remaining electron occupies the other channel. Additionally, there is a single diagram depicting all electrons with the same spin orientation.
Our research focuses on a system consisting of 3 electrons, where the electron is affiliated with the second-order special unitary group SU(2). An electron is represented by a box $\square$. Initially, by multiplying two boxes $\square\otimes\square$, we obtain the diagrams for the case of two electrons. These diagrams result in the tableaux illustrating the symmetrical case (triplet), depicted as three rows with two boxes each, and for the antisymmetrical case (singlet), the tableau associated is a column with two boxes, we obtain $$ 2\otimes 2  = 3 \oplus 1 $$.
For 3 electrons: $$ 2\otimes 2 \otimes 2 = 4 \oplus 2 \oplus 2 $$ where $\otimes$ is the direct product and $\oplus$ the direct sum. We get 4 symmetrical states, 2 symmetric mixed states, and 2 antisymmetric mixed states \cite{Taut}.

For the 4 symmetrical states, we have a Young diagram represented
by a single row (for pure antisymmetric states there would be a single column):
\ytableausetup{centertableaux}
\begin{ytableau}
1 & 2 & 3 \\
\end{ytableau}
\vspace{0.3cm}
\\
and using the notation $\ket{S,M_S}$, $S$  spin and $M_S$ spin projection, the states are:

$$\ket{\frac{3}{2},\frac{3}{2}}=\uparrow \uparrow \uparrow$$
$$\ket{\frac{3}{2},-\frac{3}{2}}=\downarrow \downarrow \downarrow$$
$$\ket{\frac{3}{2},\frac{1}{2}}=\frac{1}{\sqrt{3}}(\uparrow \downarrow \uparrow + \downarrow \uparrow \uparrow + \uparrow \uparrow \downarrow)$$
$$\ket{\frac{3}{2},-\frac{1}{2}}=\frac{1}{\sqrt{3}}(\uparrow \downarrow \downarrow + \downarrow \uparrow \downarrow + \downarrow \downarrow \uparrow)$$

To define symmetric and antisymmetric mixed states, since the spin space is degenerate, there is more than one orthogonal spin eigenstate $\Lambda_i$ for $S$ y $M_S$ set. We use the notation $\ket{S,M_S,i}$ for those states.

For the 2 symmetric mixed states, we have the tableau: \ytableausetup{centertableaux}
\begin{ytableau}
1 & 2 \\
3\\
\end{ytableau}
with states:

$$\ket{\frac{1}{2},\frac{1}{2},1}=\frac{1}{\sqrt{6}}(2\uparrow \uparrow \downarrow - \uparrow \downarrow \uparrow - \downarrow \uparrow \uparrow)$$
$$\ket{\frac{1}{2},-\frac{1}{2},1}=\frac{1}{\sqrt{6}}(2\downarrow \downarrow \uparrow - \uparrow \downarrow \downarrow - \downarrow \uparrow \downarrow)$$

For the 2 antisymmetric mixed states, we have the tableau: \ytableausetup{centertableaux}
\begin{ytableau}
1 & 3 \\
2\\
\end{ytableau}

with states: 

$$\ket{\frac{1}{2},\frac{1}{2},2}=\frac{1}{\sqrt{2}}(\uparrow \downarrow \uparrow - \downarrow \uparrow \uparrow )$$
$$\ket{\frac{1}{2},-\frac{1}{2},2}=\frac{1}{\sqrt{2}}(\uparrow \downarrow \downarrow - \downarrow \uparrow \downarrow )$$

As in Taut's work \cite{Taut}, $i=1$ labels symmetric mixed states and $i=2$ labels the antisymmetric states.

\section{Computational Details}
The ground state calculations were computed in Octopus version 10 \cite{Octopus} using the following simulation boxes: i) $\omega_0=0.5$ a.u.:  $20$~a.u.\ simulation box with $0.5$~a.u.\ spacing.
ii) $\omega_0=0.1$ a.u.: $20$~a.u.\ simulation box with $0.5$~a.u.\ spacing. iii) $\omega_0=0.05$ a.u.: $100$~a.u.\ simulation box with $0.8$~a.u.\ spacing.  iv) $\omega_0=0.005$  a.u. :  $100$~a.u.\ simulation box with $0.8$~a.u.\ spacing.

The exact calculations were done using the modelmb functionality 
in Octopus \cite{Helbig} and the code was modified to output 
the quadrupole moment in one dimension.

The interaction with the TL field is modeled as in Ref.~ \cite{ic1/ic2},
\begin{equation}
H_{I}^{l}=E_0 f(t) \sum_i^N x_i^{(l+1)}
\end{equation}
assuming dimension and position of the harmonic trap w.r.t. to the twisted light beam fulfills the requirements stated in the same Ref.
%
When $l=0$ the field couples to the dipole and only the center of mass degrees of freedom get excited (Kohn's theorem \cite{Yip}).
For the time dependence, we choose a step function 
$f(t)=\text{rect}(t-\tau/2)$, 
where $\tau=0.01 a.u.$ is the duration of the pulse.
%
%
The first-order quadrupole response to the TL probe can be computed as 
\begin{equation}
 Q^{(1)}(\omega)=FT[Q^{(1)}(t)]/FT[{\cal E}(t)]
 \label{eq:Q1}
\end{equation}
where ${\cal E}(t)=E_0 f(t)$ is the external perturbation due to the 
interaction between the harmonic trap and the TL field. 
%
%
%
%
The time-dependent quadrupole can be computed from the evolution of 
the density \mbox{$\rho(x,t)$ } as
\begin{equation}
   Q(t) = -e\langle\sum_i^{N=2}\hat{x}_i^2\rangle =
          -e\int  \rho(x,t) {\bf x}^2 dx.
\end{equation}
$Q^{(1)}(t)$ in Eq.~(\ref{eq:Q1}) is the variation of the 
quadrupole moment due to the action of a TL probe of odd topological 
charge $l$, i.e.\ $Q^{(1)}(t)= \delta Q(t)=Q(t)-Q^0$, 
with $Q^0=-e\int  \rho^0(x,t) {\bf x}^2 dx$ being the unperturbed 
quadrupole moment.
%
%
%
%
%
%
The contribution of $FT[\text{rect}(t-\tau/2)]$ is only significant around 
$\omega=0$ and will be ignored here.
The simulation was set with an value of $E_0=0.001$, $dt=0.01$ the timestep, $T=6000$ the total propagation time, all these values in a.u. The propagator used was Approximated Enforced Time-Reversal Symmetry (AETRS).

\subsection{Parity check on simulated wave function}


\begin{figure}[htbp]
   \centering
    \includegraphics[width=0.48\textwidth]{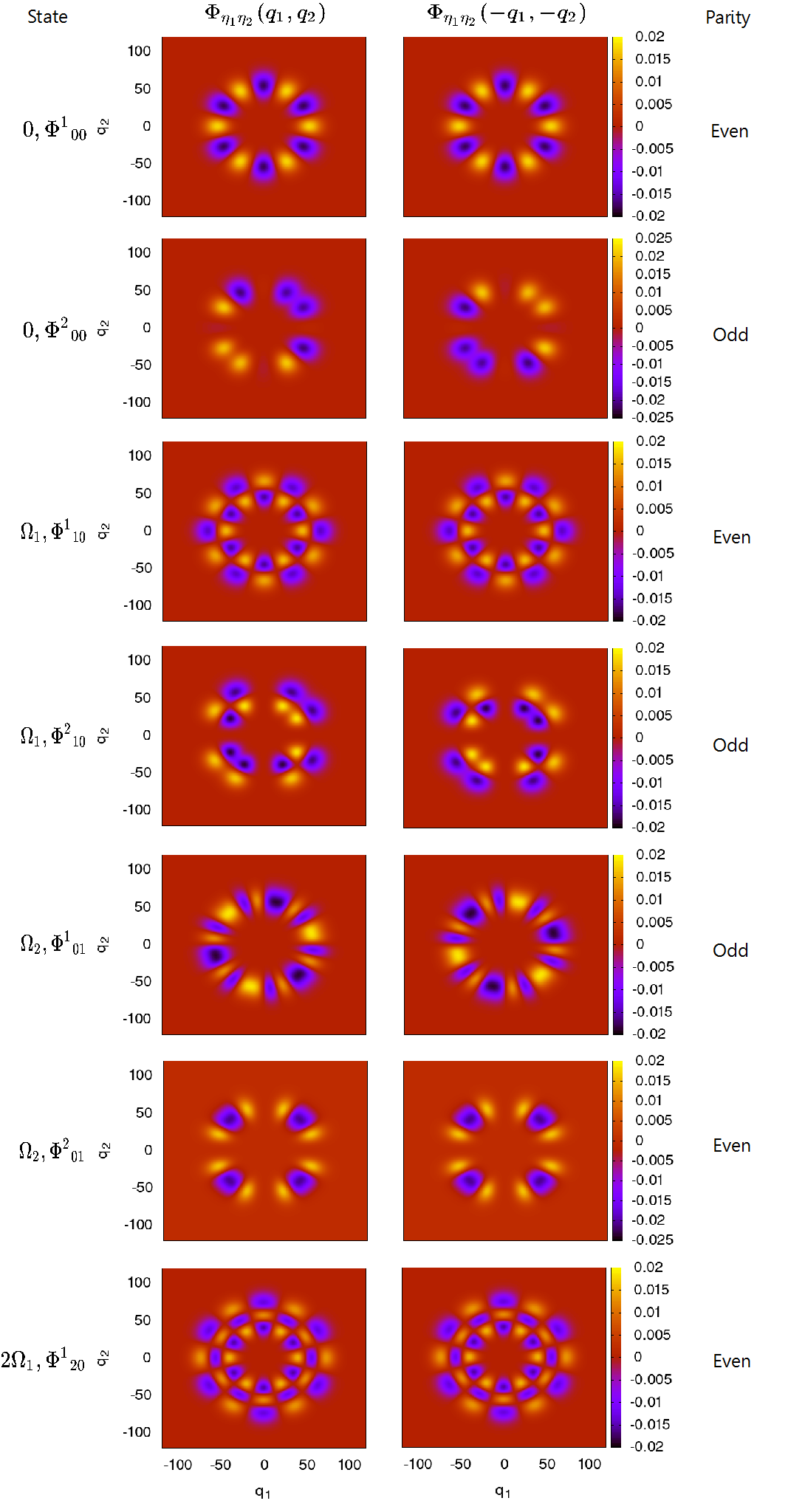}
    \caption{First 7 internal spatial eigenstates and their respective parities for $\omega_0=0.005$ a.u. (strong-correlation regime). In the strongly-correlated limit each state has degeneracy 2.}
    \label{fig:intw}
\end{figure}
To ensure the simulated wavefuction has the proper symmetry we performed a parity check. Since the Hamiltonian is separable and the center of mass wavefuction $\zeta(X)$ commutes with parity, we checked parity for the more interesting internal wavefunction $\xi(q_1,q_2)$, which needs to preserve parity as well, $\xi(-q_1,-q_2)=\pm \xi(q_1,q_2)$ \cite{Loft}. We perform the check for the most correlated case we computed, namely $\omega_0=0.005$ a.u., which is also the hardest to converge. 
In Fig.~\ref{fig:intw} we plot the wave functions $\xi(q_1,q_2)$ and $\xi(-q_1,-q_2)$ of the internal system for a harmonic trap of $\omega_0=0.005$ a .u. For each set of $\{\eta_1,\eta_2\}$, there are two wavefunctions: one even and one odd.

\end{document}